# The principle of a virtual multi-channel lock-in amplifier and its application to magnetoelectric measurement system

Jun Lu, De-An Pan and Lijie Qiao[a]

*Corrosion and Protection Center, Key laboratory of Environmental Fracture (Ministry of Education), University of Science and Technology Beijing, 100083, P. R. China*

This letter presents principles and applications of a virtual multi-channel lock-in amplifier that is a simple but effective method to recover small ac signal from noise with high presison. The fundamentals of this method are based on calculation of cross-correlation function. Via this method, we successfully built up a magnetoelectric measurement system which can perform precise and versatile measurements without any analog lock-in amplifier. Using the virtual multi-channel lock-in amplifier, the output of the magnetoelectric measurement system is extensively rich in magnetoelectric coupling behaviors, including coupling strength and phase lag, under various dc bias magnetic field and ac magnetic field.

## I. Introduction

Lock-in amplifier is widely used to measure a physical quantity buried in noise because it can single out the component of the signal at a specific reference frequency and filter out noise signals at frequencies other than the reference frequency using a technique known as phase sensitive detection.[1] However, most commercial analog or digital lock-in amplifiers still have some disadvantages that are (i) single signal processing unit in one instrument if not specifically customized; (ii) low frequency limit in the magnitude of 100 kHz for most of them; and (iii) high cost. Subsequently, it must be a formidable task to measure a system where two or more noisy signals need to be analyzed at the same time and where the frequency of interest is higher than the frequency limit of lock-in amplifiers at hand. Computer-based virtual lock-in amplifiers was applied by some authors,[2] but so far no lock-in amplifier has been realized yet to overcome all of the these drawbacks.

[a] Corresponding author. Fax: +86 10 6233 2345
 *E-mail address*: lqiao@ustb.edu.cn

Here we have designed a computer-based multi-channel virtual lock-in amplifier using principle of correlation analysis, which is proved to defeat these imperfections without complicated design. We intend to apply this method to the measurement of magnetoelectrics, which has increasing promise for science and technology.[3] Magnetoelectric coefficient $\alpha_{ME} = dE/dH = dV/(B \cdot dH)$ is the most critical indicator for the magnetoelectric coupling properties in this kind of materials, where $V$ is the induced magnetoelectric voltage, $H$ is the exciting ac magnetic field and $B$ is the thickness that $V$ across the laminate.[4] Practically, d$H$ is produced by small an ac magnetic field superimposed onto dc bias magnetic fields in the same direction, so d$V$ is also an ac signal. Herein d$V$ and d$H$ should be dynamically observed synchronously but unfortunately, in most cases they were recorded separately due to the above-mentioned weakness of commercial lock-in amplifiers, which leaded errors especially when the measurement frequencies increase and both signals become quite noisy.[4-5] This letter presents the result that such a problem can be solved properly by technique of virtual multi-channel lock-in amplifier shown in figure 1.

We first develop this method in Sec. II, and in Sec. III we show, with specific examples, how the method allows implementation of automatically and precise magnetoelectric assessment with rich output.

## II. Apparatus and Principle

As shown in figure 1, the virtual multi-channel lock-in amplifier could be divided into three parts that are (i) a front-end sampling device, typically an oscilloscope, to dynamically and synchronically convert analog multi-channel signals into digital series and transfer them into computer; (ii) a virtual inner reference, embedded in computer, to remote control signal generator and send unit sinusoidal reference into correlation analyzer where the frequencies are completely consistent in both paths; and (iii) a virtual multi-channel correlation analyzer, also embedded in computer, to calculator amplitudes and phases of analog multi-channel signals, which is the central device of the virtual lock-in amplifier.

In figure 1, sample 1 is put into a Helmholtz coil 2, and in the middle of two electromagnets 3. The dc biased magnetic field produced by the magnets 3 is close-loop controlled by computer via interface of AD/DA cards, which read strength of dc magnetic field from gaussmeter 4 and converted control signal into analog voltages to tune output of the dc source 5.

Helmholtz coil will produce a small ac magnetic field, which is supplied by the power amplifier 6 and whose frequency is controlled by the signal generator 7. The signal generator is driven by the inner virtual reference 8, so the frequency of the small ac magnetic field is originally handled by the inner virtual reference.

The sample under test is leaded out by coax to shield noise. The measured signal produced by the sample 1 is input into the prediffratiate device 9 to reject common mode signals. Finally, the magnetoelectric signal is sampled by the oscilloscope 10 then transfers into the correlation analyzer 11 in the computer, synchronized with the ac signal which indicated the intensity of the small ac magnetic field.

In the correlation analyzer, the measured signal from the oscilloscope and the reference signal from the virtual reference 8 will be synchronously analyzed and the results are input into the recorder unit 12.

The principle of the correlation analyzer is as follows. Here consider the situation where the data of interest are measurements of two continuous random processes $\{x(t)\}$ and $\{y(t)\}$. When the time delay between $x(t)$ and $y(t)$ equals to zero, the cross-correlation function between $x(t)$ and $y(t)$ has an explicit form $R_{xy} = \lim_{T \to \infty} \frac{1}{T} \int_0^T x(t)y(t)dt$ and particularly in digital case $R_{xy} = \frac{1}{N} \sum_{k=1}^{N} [x(k)y(k)]$, where N is the total number of the series.[6] For a stationary random time history of the form $V_m(t) = s(t) + n(t)$, where $n(t)$ is Gaussian random noise and $s(t) = A\sin(\omega t + \phi)$ is a sinusoidal signal. To obtain $\phi$ and $A$ of the signal, we introduced a pair of reference signals $V_r(t) = \sin(\omega t)$ and $V_{r'}(t) = \sin(\omega t + \pi/2)$. Because a sinusoidal reference has no correlation with random noise, i.e., $R_{nr} = R_{nr'} = 0$, the correlation function between $V_m$ and $V_r$ ($V_{r'}$) is equal to $R_{mr} = \frac{1}{N} \sum_{k=1}^{N} [V_m(k)V_r(k)]$ ($R_{mr'} = \frac{1}{N} \sum_{k=1}^{N} [V_m(k)V_{r'}(k)]$). As a result, the phase and the amplitude of the sinusoidal signal can be calculated in the forms $\phi = \arctan(R_{mr'}/R_{mr})$ and $A = 2R_{mr}/\cos\phi$, respectively.[7]

When multi-channel sinusoidal signals with the same frequency but different phase need to be measured, we perform correlation calculation for each signal with the same pair of references and estimate amplitude and phase for each of them. In the case of magnetoelectric measurement, we can obtain $A_{dV}$ and $\phi_{dV}$ for induced ac magnetoelectric voltage and $A_{dH}$ and $\phi_{dH}$ for ac magnetic field. Thus phase lag of magnetoelectric coupling $\phi_{ME} = \phi_{dV} - \phi_{dH}$ is acquired, as well as $|\alpha_{ME}| = A_{dV}/A_{dH}$. Note that it is not difficult to understand that the phase lag $\phi_{ME}$ is independent on phase of references.

## III. Experimental verification

In this section, we provide examples to verify the method derived in the previous sections. The sample under test is a Ni/PZT/Ni trilayer, where PZT, a typical piezoelectrics, was molded and sliced to a square with dimension of 25x25x0.8 mm$^3$ and Ni was electrodeposited on both side with the total thickness about 2x0.4 mm, which acts as piezomagnetic function. When both dc bias magnetic field and ac magnetic field are applied along the longitudinal axis of the laminate and the Ni electrodes are perpendicular to the normal of the laminate, the $\alpha_{ME}$ is notated by $\alpha_{ME,31}$, where the subscript "3" and "1" indicate the direction of the magnetic field induced electric field and the magnetization direction of Ni layers, respectively.

At the beginning of the measurement, the $|\alpha_{ME,31}|$ dependence on dc bias magnetic field $H_D$ was observed with the frequency of ac magnetic field equal to 1 kHz. Figure 2 shows that with the rise of $H_D$, $|\alpha_{ME,31}|$ increases first, reaches a maximum at $H_D = 150$ Oe, then decreases slowly. The trend of $|\alpha_{ME,31}|$ versus bias magnetic field in figure 2 can be interpreted by the magnetostriction characterization of Ni which saturate at a particular magnetic field leading to the maximum of $|\alpha_{ME,31}|$ at about $H_D$ = 150 Oe.[8]

Under the constant dc bias magnetic field $H_D$ = 150 Oe, the dependence of $|\alpha_{ME,31}|$ and phase lag on frequency of ac magnetic field was measured. Figure 3 shows that a sharp resonance peak takes place at about 86 kHz which is associated with the electromechanical resonance in the primary transverse mode, and there are another two smaller peaks at about 59 kHz and 73 kHz, respectively, indicating electromechanical resonance in other modes.[9] The primary peak at 86 kHz has been highlighted in figure 4 compared with data measured by an analog lock-in amplifier (EG&G PARC Model 124A from Princeton Applied Research), where the ac magnetoelectric voltage d$V$ and the intensity of the ac magnetic field d$H$ were measured separately for any particular frequency. Figure 4 demonstrates the high precision of our design because the amplitudes of magnetoelectric coupling measured by our virtual lock-in amplifier are quite consistent with that measured by the analog counterpart.

To extensively investigate correlation between magnetoelectric coupling of the Ni/PZT/Ni trilayer and frequency of the ac magnetic field combined with dc bias magnetic field, the close loop control module of dc bias magnetic field was imposed into the measurement system. Figure 5 and figure 6 illustrate the resonance phase diagram of magnetoelectric coupling $|\alpha_{ME,31}|$ and phase lag under dc bias magnetic field, where the resonance frequency keeps invariant no matter what the dc bias magnetic field is. The interface at $H_D$ = 0 Oe in figure 5 and figure 6 indicate same magnetoelectric coupling behaviors except a sign change in both sides, which can also be explained by the magnetostriction versus external dc magnetic field $H_D$ of the Ni piezomagnetic layers. Because magnetostriction versus $H_D$ is an even function and saturate at certain magnetic field, its differential function and then magnetostriction induced magnetoelectric coefficient $\alpha_{ME,31}$ versus dc bias magnetic field have positive and negative maximum at positive and negative side of dc magnetic field, respectively, resulting in the sign change along axis of dc bias magnetic field in figure 6.[8]

## IV. Summary

The benefits of the virtual multi-channel lock-in amplifier have been clearly verified where the rich data were precisely recovered from the multi-channel small noisy signals. The implication of this method let us envision low-cost effective solutions to defeat noise in wider fields such as scientific measurements, engineering and medical care.

**Figure Captions Page**

FIG. 1. Schematic diagram of the virtual multi-channel lock-in amplifier for magnetoelectric measurement.

FIG. 2. The magnetoelectric coefficient $|\alpha_{ME,31}|$ dependence on dc bias magnetic field at 1 kHz ac magnetic field.

FIG. 3. The magnetoelectric coefficient $|\alpha_{ME,31}|$ and phase lag ($\phi_V$-$\phi_H$) versus frequency $f$ of ac magnetic field with the constant dc bias magnetic field $H_D$ = 150 Oe.

FIG. 4. The magnetoelectric coefficient $|\alpha_{ME,31}|$ versus frequency $f$ of ac magnetic field measured by the virtual lock-in amplifier and an analog one with the constant dc bias magnetic field $H_D$ = 150 Oe.

FIG. 5. The 3-D mapping of $|\alpha_{ME,31}|$ (magnetoelectric coefficient) versus $f$ (frequency of ac magnetic field) and $H_D$ (dc bias magnetic field).

FIG. 6. The 3-D mapping of $\phi_V$-$\phi_H$ (magnetoelectric coupling phase lag) versus $f$ (frequency of ac magnetic field) and $H_D$ (dc bias magnetic field).

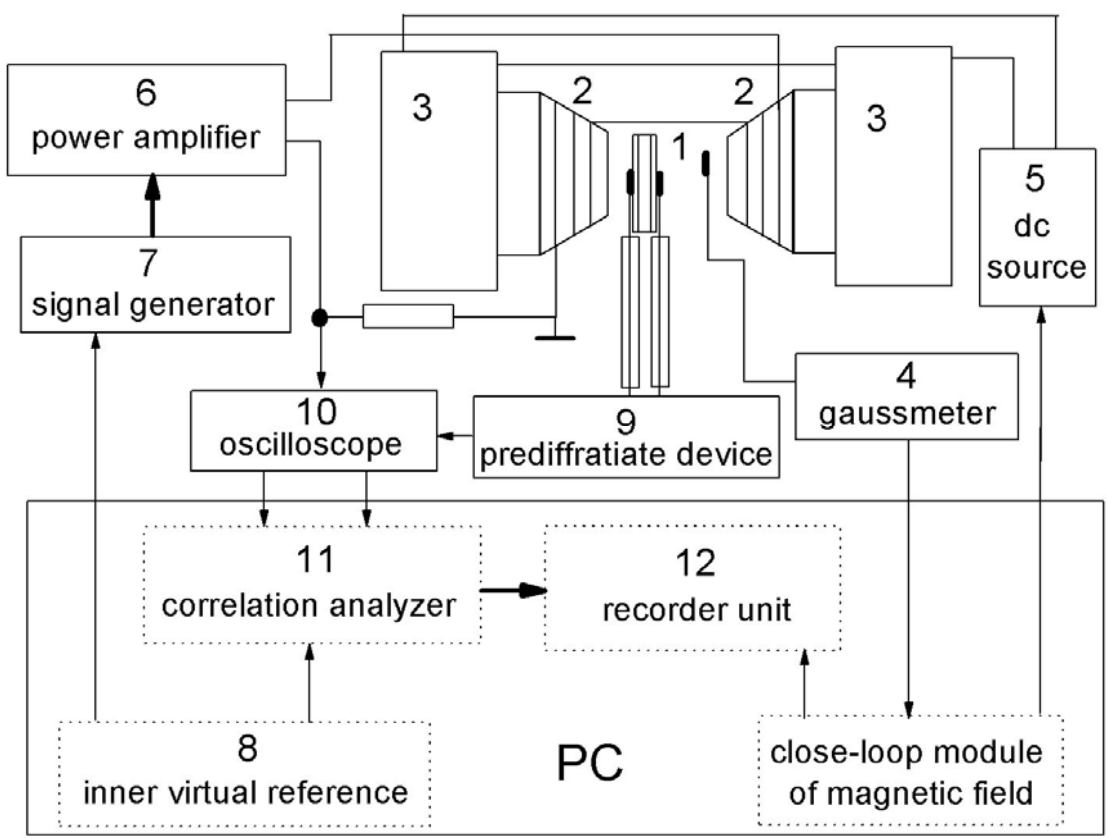

FIG. 1. Schematic diagram of the virtual multi-channel lock-in amplifier for magnetoelectric measurement.

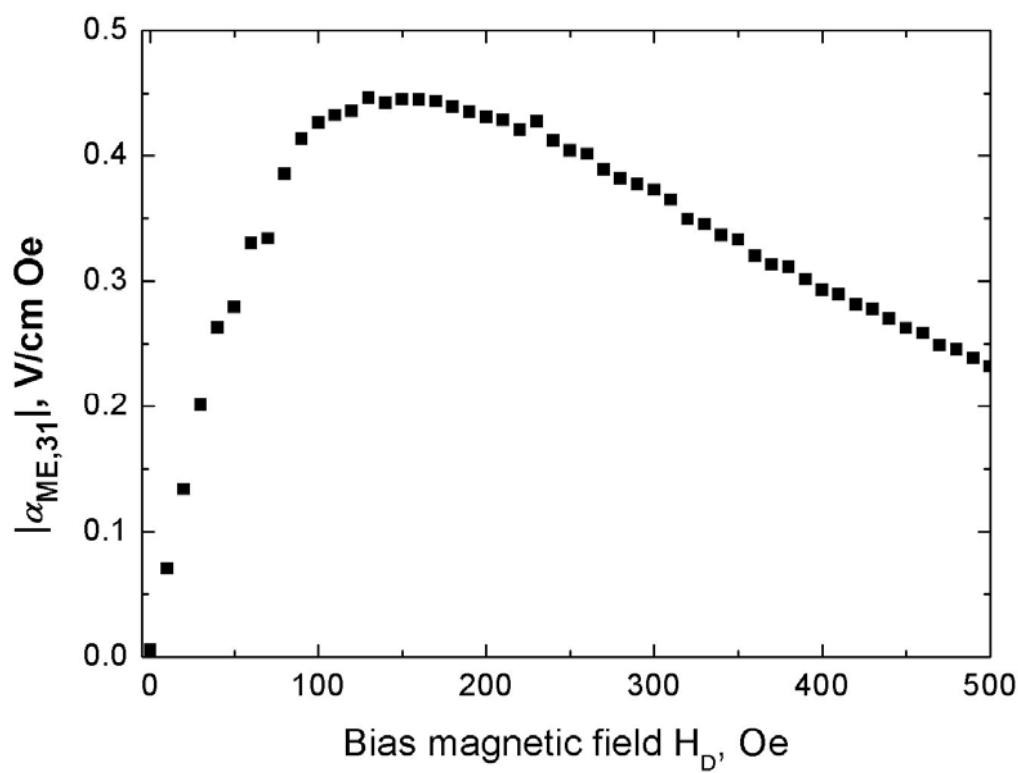

FIG. 2. The magnetoelectric coefficient $|\alpha_{ME,31}|$ dependence on dc bias magnetic field at 1 kHz ac magnetic field.

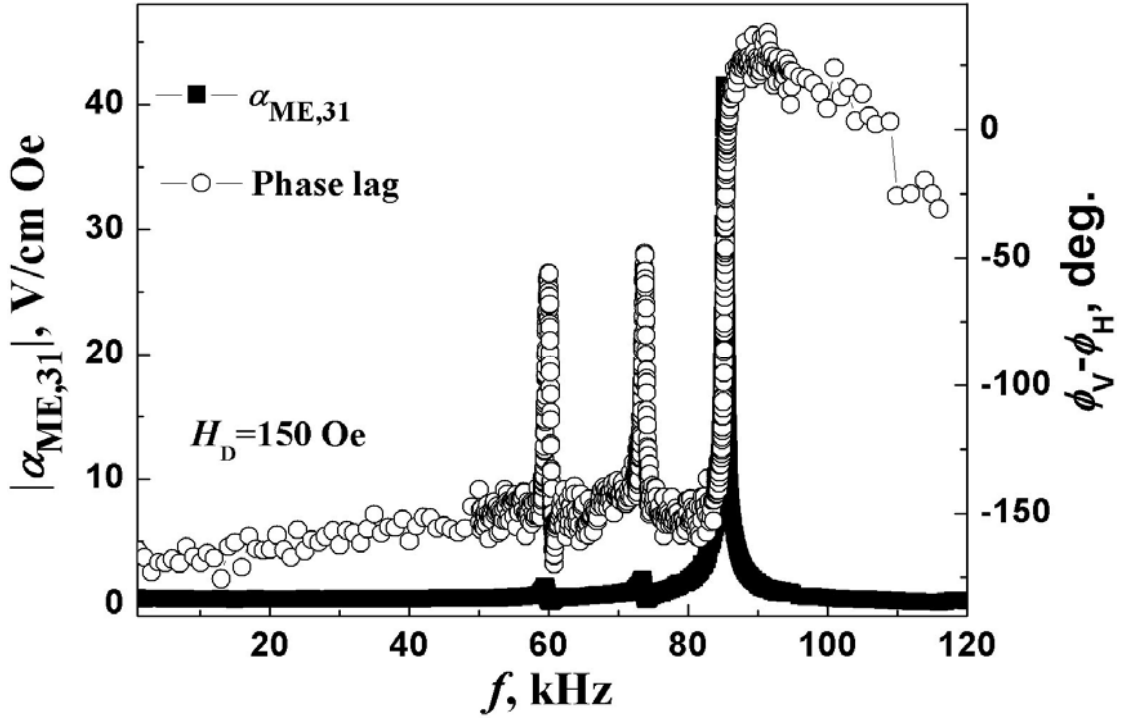

FIG. 3. The magnetoelectric coefficient $|\alpha_{ME,31}|$ and phase lag ($\phi_V - \phi_H$) versus frequency $f$ of ac magnetic field with the constant dc bias magnetic field $H_D$ = 150 Oe.

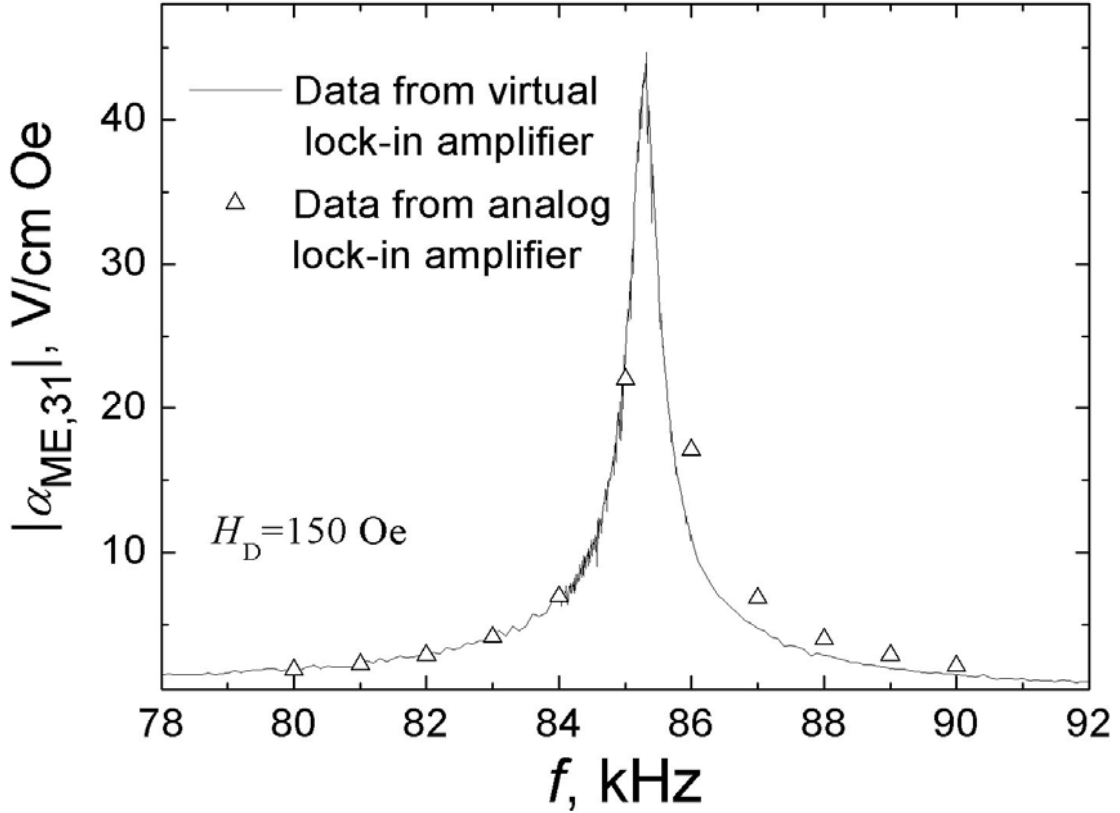

FIG. 4. The magnetoelectric coefficient $|\alpha_{ME,31}|$ versus frequency $f$ of ac magnetic field measured by the virtual lock-in amplifier and an analog one with the constant dc bias magnetic field $H_D$ = 150 Oe.

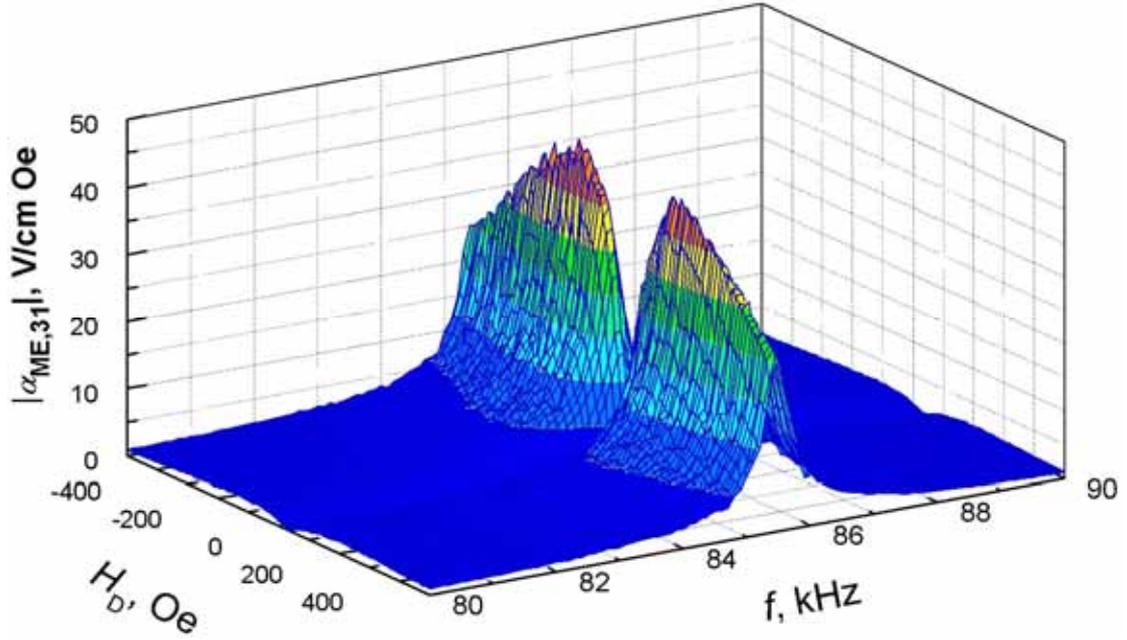

FIG. 5. The 3-D mapping of $|\alpha_{ME,31}|$ (magnetoelectric coefficient) versus $f$ (frequency of ac magnetic field) and $H_D$ (dc bias magnetic field).

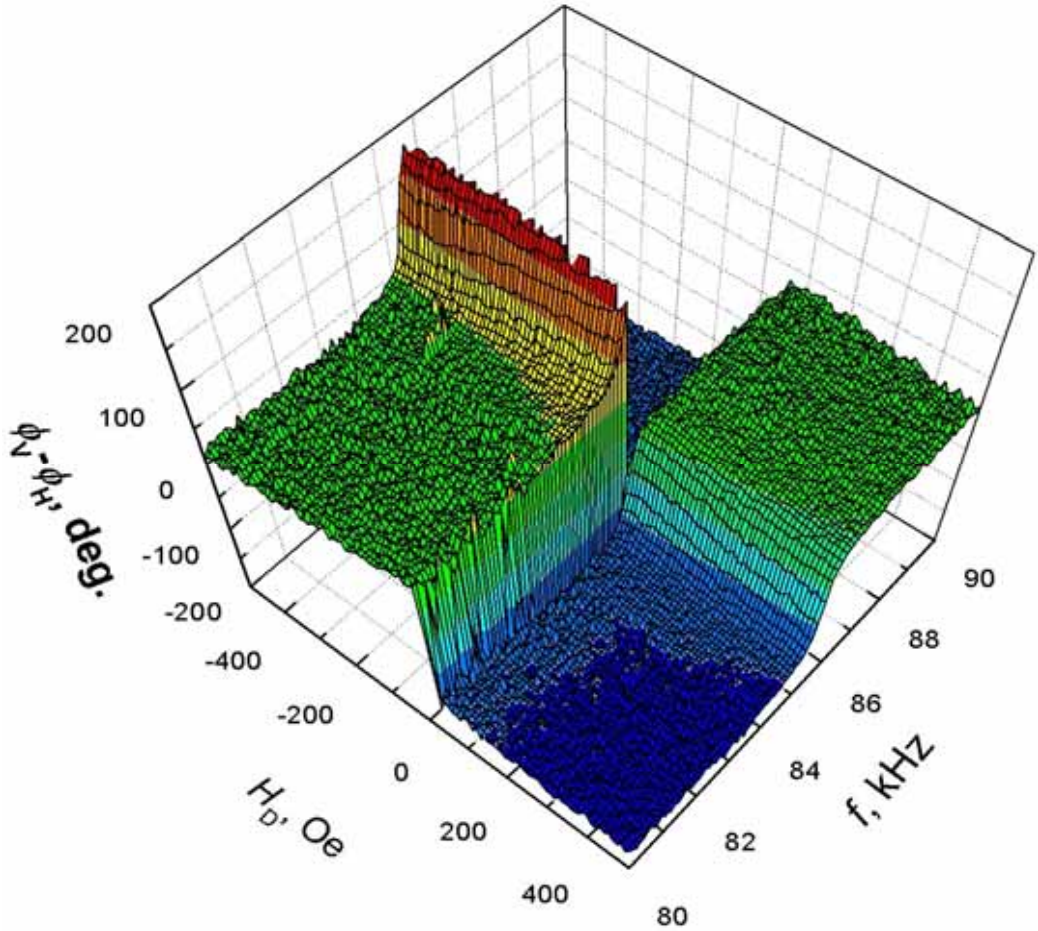

FIG. 6. The 3-D mapping of $\phi_V$-$\phi_H$ (magnetoelectric coupling phase lag) versus $f$ (frequency of ac magnetic field) and $H_D$ (dc bias magnetic field).